\documentclass{elsart}
\usepackage{epsfig}
\usepackage{graphics}
\usepackage{color}
\definecolor{mygreen}{rgb}{0,.8,0}
  \def\black{\color{black}}

\def\ifm#1{\relax\ifmmode#1\else$#1$\fi}
\def\DAF{DA\char8NE}  
\def\f{\ifm{\phi}}  
\def\ab{\ifm{\sim}}  \def\x{\ifm{\times}}
\def\gam{\ifm{\gamma}}  \def\pic{\ifm{\pi^+\pi^-}}
\def\pt#1,#2,{\ifm{#1\x10^{#2}}}  \def\dif{\ifm{{\rm d}\,}}
\renewcommand{\to}{\ensuremath{\rightarrow}}
\def\ao{\ifm{a_0}}  \def\po{\ifm{\pi^0}}  \def\et{\ifm{\eta}}

\makeatletter
\newdimen\z@ \z@=0pt 
\newskip\z@skip \z@skip=0pt plus0pt minus0pt
\def\m@th{\mathsurround=\z@}
\def\ialign{\everycr{}\tabskip\z@skip\halign} 
\def\eqalign#1{\null\,\vcenter{\openup\jot\m@th
  \ialign{\strut\hfil$\displaystyle{##}$&$\displaystyle{{}##}$\hfil
      \crcr#1\crcr}}\,}
\makeatother

\newcommand{\aff}[2]{Dipartimento di Fisica dell'Universit\`a #1 e Sezione INFN, #2, Italy.}
\newcommand{\affd}[1]{Dipartimento di Fisica dell'Universit\`a e Sezione INFN, #1, Italy.}
\newcommand{\kl}{\ensuremath{K_L}}
\newcommand{\ks}{\ensuremath{K_S}}

\newcommand{\GeV}{\ensuremath{\mathrm{GeV}}}

\newcommand{\ps}{\ensuremath{\mathrm{ps}}}
\newcommand{\um}{\ensuremath{\mu\mathrm{m}}}
\newcommand{\mm}{\ensuremath{\mathrm{mm}}}
\begin{document}

\begin{frontmatter}



\title{Study of the Decay \f\to\et\po\gam\ with the KLOE Detector}

\collab{The KLOE Collaboration}
\author[Na] {A.~Aloisio},
\author[Na]{F.~Ambrosino},
\author[Frascati]{A.~Antonelli},
\author[Frascati]{M.~Antonelli},
\author[Roma3]{C.~Bacci},
\author[Frascati]{G.~Bencivenni},
\author[Frascati]{S.~Bertolucci},
\author[Roma1]{C.~Bini},
\author[Frascati]{C.~Bloise},
\author[Roma1]{V.~Bocci},
% ...

\address[Bari]{\affd{Bari}}
\address[Frascati]{Laboratori Nazionali di Frascati dell'INFN, 
Frascati, Italy.}
\address[Karlsruhe]{Institut f\"ur Experimentelle Kernphysik, 
Universit\"at Karlsruhe, Germany.}
\address[Lecce]{\affd{Lecce}}
\address[Na]{Dipartimento di Scienze Fisiche dell'Universit\`a 
``Federico II'' e Sezione INFN,
Napoli, Italy}
\address[Pisa]{\affd{Pisa}}
\address[Energ]{Dipartimento di Energetica dell'Universit\`a 
``La Sapienza'', Roma, Italy.}
\address[Roma1]{\aff{``La Sapienza''}{Roma}}
\address[Roma2]{\aff{``Tor Vergata''}{Roma}}
\address[Roma3]{\aff{``Roma Tre''}{Roma}}
\address[StonyBrook]{Physics Department, State University of New 
York at Stony Brook, USA.}
\address[Trieste]{\affd{Trieste}}
\address[Virginia]{Physics Department, University of Virginia, USA.}
\address[Beijing]{Permanent address: Institute of High Energy 
Physics, CAS,  Beijing, China.}
\address[Moscow]{Permanent address: Institute for Theoretical 
and Experimental Physics, Moscow, Russia.}

\begin{abstract}
In a sample of \pt5.3,7, \f-decays observed with the KLOE detector at the Frascati
\f-factory \DAF\ we 
find 605 \et\po\gam\ events with $\eta\rightarrow\gamma\gamma$
and 197 \et\po\gam\ events with $\eta\rightarrow\pi^+\pi^-\pi^0$. 
The decay \et\po\gam\ is dominated by the process 
\f\to\ao\gam. From 
a fit to the \et\po\ mass spectrum we find BR(\f\to\ao(980)\gam)=
$(7.4\pm0.7)\times 10^{-5}$.
\end{abstract}
\end{frontmatter}

There is no clear understanding of the \ao(980) and $f_0$(980) mesons 
in the quark model. It has been argued that these mesons might not in fact 
be $q\bar q$ states but rather 4-quark states ($q\bar qq\bar q$)
\cite{4quarks} or 
$K\overline K$ molecules \cite{kk}. The amplitude for the E1 transition \f\to\ao\gam\ 
to a state $a_0$ with $J^P(a_0)=0^+$ is proportional to $k\x f(m^2)$, 
with $k$ the momentum and $m$ the mass of the $a_0$. The differential 
decay width $\dif\Gamma/\dif m$ for \f\to\ao\gam\ has the form 
$k^2\x F(m^2)$, where $F$ is a Lorentzian for the $a_0$ 
times a damping factor depending on the \f\ and $a_0$ wave functions. 
The shape of the \ao\ signal in \f\ decay is therefore grossly distorted, 
acquiring a large tail at low $m$. 

According to different interpretations the 
$\phi\rightarrow a_0\gamma$ branching ratio can range from 10$^{-5}$ for 
$q\bar q$ and $K\overline K$ to 10$^{-4}$ for $q\bar qq\bar 
q$~\cite{Achasovold}.
The ratio BR(\f\to$f_0$\gam)/BR(\f\to\ao\gam) also depends on the 
structure of the scalars~\cite{close}. Chiral perturbation 
theory also attempts to estimate the amplitude 
and the distortion of the \ao\ line shape observed in \f\ meson 
decays~\cite{chPT}. Production of the \ao\ meson, followed by \ao\to\et\po\ 
dominates the final state \et\po\gam\ in \f-decays. A small 
contribution is due to \f\to$\rho^0$\po, $\rho^0$\to\et\gam.

We present in the following a study of the decay \f\to\et\po\gam\ performed 
with the KLOE detector~\cite{Kloep} at the Frascati 
\f-factory~\cite{Dafneref} \DAF. \DAF\ delivered about 16 pb$^{-1}$ 
during the year 2000, operating at a total energy of 1020 MeV. 
About \pt5.3,7, \f\ mesons were produced and collected by KLOE.
We searched for two types of events. Type 1 events have five photons
from \f\to\et\po\gam~ with \et\to\gam\gam. Type 2 
events have five photons plus two charged tracks corresponding 
to the same initial decay but with \et\to\pic\po. Observation 
of type 1 events has been reported by SND~\cite{snda0} and CMD-2~\cite{cmd2a0}
experiments 
at Novosibirsk. In this work we report the first observation 
of type 2 events. While the yield for type 2 events is lower, 
there is no physical background.

The KLOE detector consists of a large cylindrical drift chamber 
surrounded by a lead-scintillating fiber electromagnetic 
calorimeter. A superconducting coil provides a 0.52 T field. 
The drift chamber~\cite{DCNim}, 4~m in diameter and 3.3~m
long, has 12,582 all-stereo tungsten sense wires and 37,746 aluminum field 
wires.
The chamber shell is made of carbon fiber-epoxy composite and the gas used is a
90\% helium, 10\% isobutane mixture.
These features maximize transparency to photons and reduce $\kl\to\ks$ regeneration
and multiple scattering. The position resolutions are 
$\approx 150~\um$ in the coordinate transverse to the wire direction and
$\approx 2~\mm$ in the longitudinal one. The momentum resolution is
$\sigma(p_{\perp})/p_{\perp}\approx 0.4\%$. Vertices are reconstructed with a
spatial resolution of $\approx 3~\mm$.
The calorimeter~\cite{EmCnim} is
divided into a barrel and two end-caps, for a total of 88 modules, and covers
98\% of the solid angle. The modules are read out at both ends by photomultipliers;
the readout granularity is \ab4.4\x4.4~$cm^2$, for a total of
2,440 ``cells.'' The arrival times of particles and the positions
in three dimensions
of the energy deposits are determined from the signals at the two ends.
Cells close in time and space are grouped into a calorimeter cluster. The cluster
energy $E_{\mathrm{CL}}$ is the sum of the cell energies.
The cluster time $t_{\mathrm{CL}}$ and position $\vec{r}_{\mathrm{CL}}$ are
energy weighed averages. Time and energy resolutions are $\sigma_E/E =
5.7\%/\sqrt{E (\GeV)}$ and  $\sigma_t = 57~\ps/\sqrt{E (\GeV)}\oplus50~\ps$,
respectively; space resolution is $\approx 1~cm$ in all the three coordinates. 
The detector trigger~\cite{TRGnim} uses calorimeter and chamber information.

We first consider the case of $\phi\to\eta\pi^0\gamma$ decays in which 
$\eta\to\gamma\gamma$.
These events are characterized by five prompt photons.
A photon is detected as a calorimeter cluster with
$E_{\mathrm{CL}} >$ 3 MeV, with no track pointing to it, and satisfying
$|t_{\mathrm{CL}}-|\vec{r}_{\mathrm{CL}}|/c|<5\sigma_t(E_{\mathrm{CL}})$. 
Photons with $|$cos$\theta|>0.93$ are rejected to avoid machine background. 
Exactly five prompt photons are required.
The main background processes are:
\begin{enumerate}
  \item $\phi\to\pi^0\pi^0\gamma$ dominated by \f\to$f_0$\gam ;
  \label{ppg}
  \item $e^+e^-\to\omega\pi^0$ with
    $\omega\to\pi^0\gamma$
  \label{wpn}
  \item \f\to\et\gam\ with \et\to\gam\gam
  \label{eg3}
  \item \f\to\et\gam\ with $\eta\to\pi^0\pi^0\pi^0 $
  \label{eg7}
\end{enumerate}
Process (3) can mimic five photon events due to energy cluster splitting or
accidental background while in process (4) photons can escape detection.
Background from \f\to\ks\kl\ with \ks\to$\pi^0\pi^0$ and \kl\ interacting
in the calorimeter is negligible after requiring the sum of
the energy of the five prompt photons to be greater than 
700 MeV.
About 1.5$\times 10^4$ events survive cuts.
A first kinematic fit, in which
4-momentum conservation and $t-|\vec{r}|/c=0$
for each prompt photon are required, is performed.
Background from processes (3) and (4) is reduced by
requiring $\chi^2$/ndf$<$3.
Events with a $\gamma\gamma$ pair having an invariant mass close to the $\eta$
mass within 30 MeV are retained. Then events with one of the three remaining
photons having energy greater than 340 MeV are rejected to reduce background 
(3);
$2.5\times 10^3$ events remain after cuts.

\begin{figure}
  \begin{center}
    \mbox{\epsfig{file=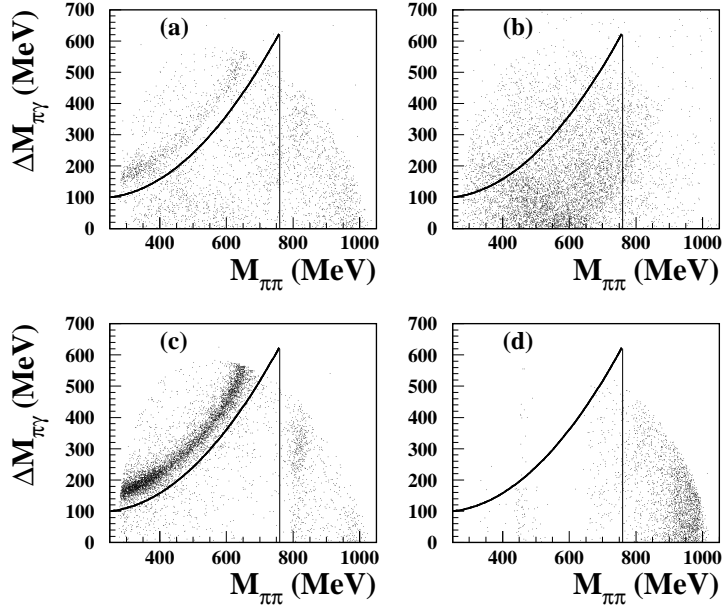,height=.4\textheight}}
    \caption{Invariant mass difference between the two $\pi^0\gamma$
      combinations versus the $\pi^0\pi^0$ mass in hypothesis (ii) (see
      text); (a) data, (b) MC $\phi\to\eta\pi^0\gamma$, (c)
      MC $e^+e^-\to\omega\pi^0\to\pi^0\pi^0\gamma$, (d)
      MC $\phi\to\pi^0\pi^0\gamma$. The solid lines delimit the selected 
      region.}
    \label{dalitz}
  \end{center}
\end{figure}
\begin{figure}
  \begin{center}
    \begin{tabular}{ c c }
\mbox{\epsfig{file=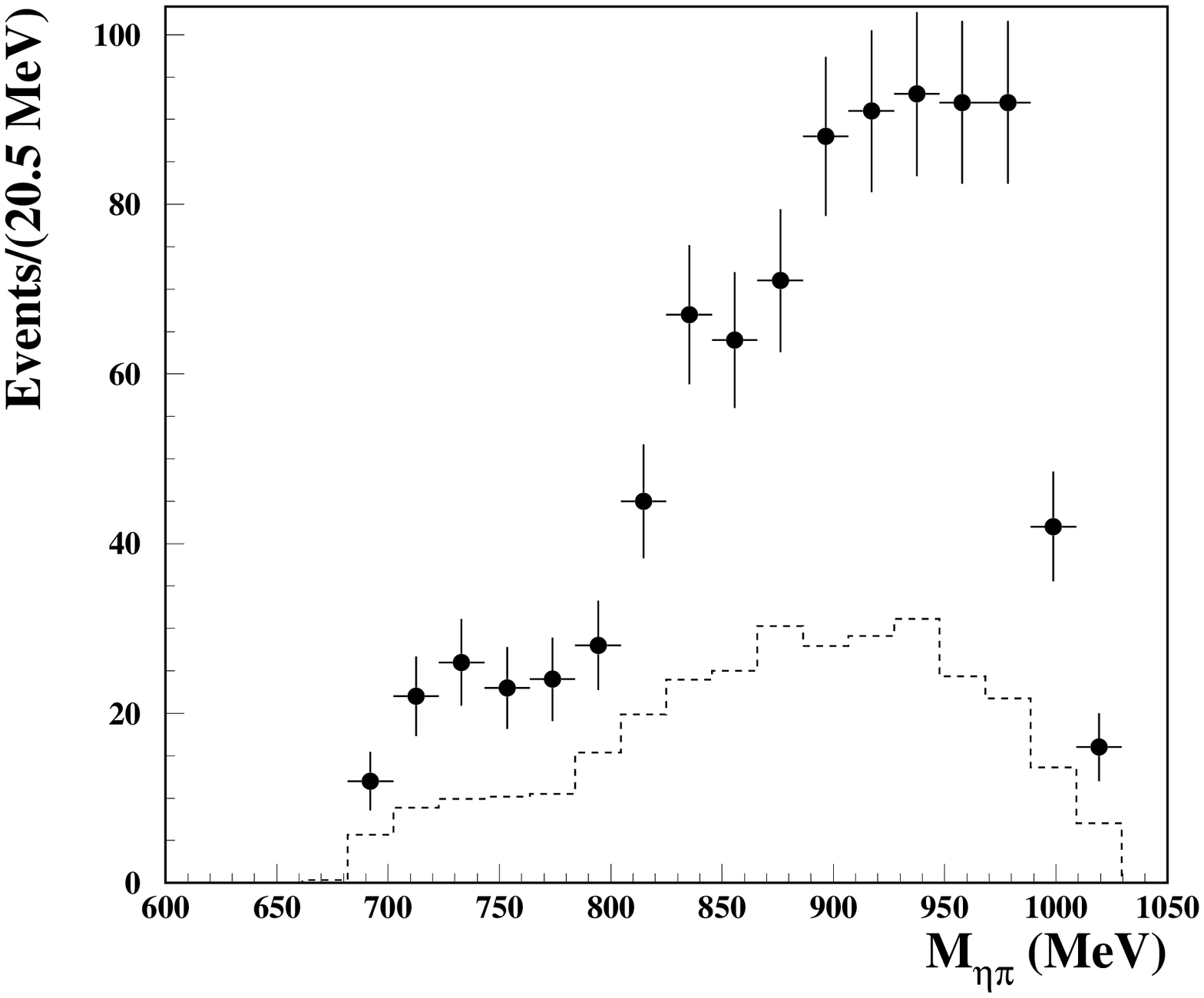,height=.255\textheight}} &
\mbox{\epsfig{file=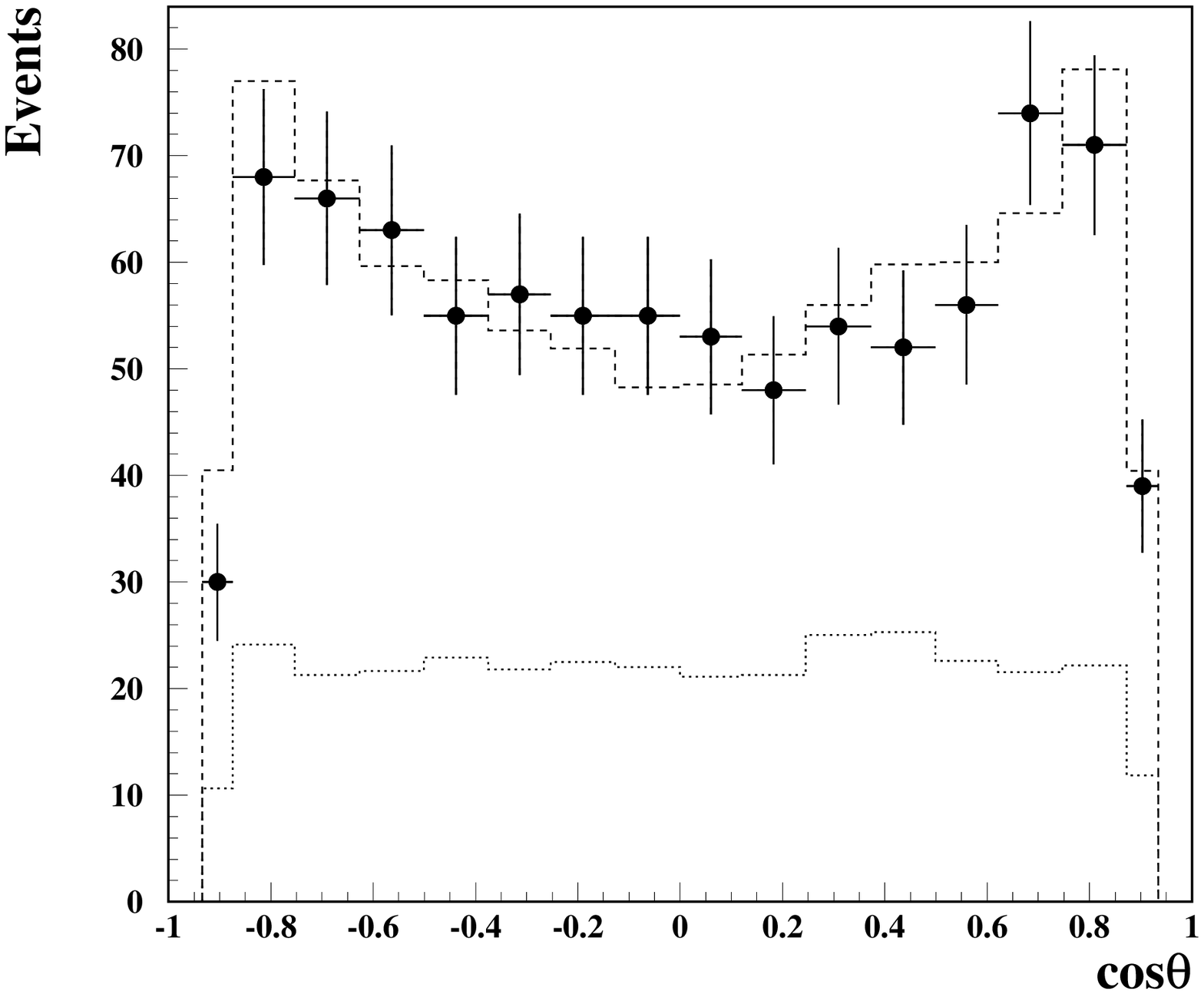,height=.255\textheight}} \\
   \end{tabular}
    \caption{$\eta\to\gamma\gamma$ sample. 
      Left: $\eta\pi^0$ invariant mass spectrum (points),
      residual background contribution (dashed histogram); Right:
      cos$\theta_{\gamma}$ distribution of the unassociated photon, comparison
      between data (points) and MC signal (dashed histogram) and MC background
      (dotted histogram).}
    \label{spectrum}
  \end{center}
\end{figure}
\begin{figure}
  \begin{center}
    \mbox{\epsfig{file=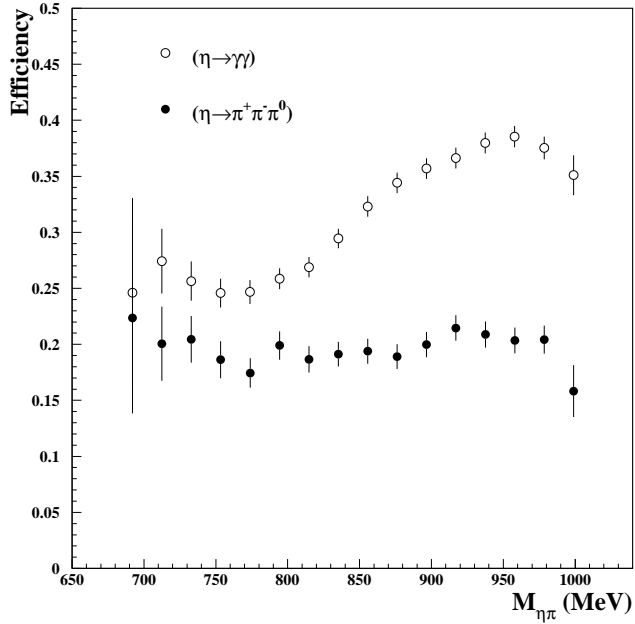,height=.4\textheight}}
    \caption{ Selection efficiency as a function of $M_{\eta\pi}$ for type 1
      and type 2 final states.}
    \label{effvsm}
  \end{center}
\end{figure}

The best photon pairing is found by matching the invariant masses of the
$\gamma\gamma$ pairs to the intermediate particles masses, either (i) 
$\eta$ and $\pi^0$ or (ii) 2 $\pi^0$.
A second kinematic fit is then performed, with constraints on the masses of the
intermediate particles, in both hypotheses.
For hypothesis (i) we retain events with $\chi^2$/ndf$<$3.
The sample is still dominated by $\pi^0\pi^0\gamma$ events from
processes (1) and (2).
These two contributions can be discriminated from the signal by exploiting
the result of the fit with
hypothesis (ii). For each of the two $\pi^0\gamma$ combinations,
an invariant mass $M_{\pi\gamma}$ is obtained;
in Fig. \ref{dalitz} the absolute value of their difference $\Delta
M_{\pi\gamma}$
is plotted versus the $\pi^0\pi^0$
invariant mass $M_{\pi\pi}$, both for data and Montecarlo (MC) events.
The $\omega\pi^0$ contribution is represented by the curved band, while
 $\phi\to\pi^0\pi^0\gamma$ events are mainly located at high 
 values of $M_{\pi\pi}$.
In order to select a clean $\phi\to\eta\pi^0\gamma$ sample, the region below
the solid curve of Fig. \ref{dalitz} and with $M_{\pi\pi}<760~MeV$ 
is retained.
The final sample consists of 916 events. The spectrum of the 
$\eta\pi^0$ invariant mass 
$M_{\eta\pi}$
is shown in Fig. \ref{spectrum} together with the expected
distribution for the background. In Fig. \ref{spectrum} the distribution of
cos$\theta_{\gamma}$ of the unassociated photon is also shown, exhibiting
the expected 1+cos$^2\theta_{\gamma}$ behaviour.
The efficiency for the identification of signal events 
is evaluated by applying the whole
analysis chain to a sample
of simulated $\phi\to\eta\pi^0\gamma$ events generated with a uniform
$M_{\eta\pi}$ distribution. The selection efficiency as a function of
$M_{\eta\pi}$ is shown in Fig. \ref{effvsm}. The average over the
whole mass spectrum is 32.4\%.
The background rejection factors and the expected number of events, 
as obtained by MC simulation, are given
in Tab. \ref{effi}.
The total number is 309$\pm$20 background events.
\begin{table}[hb]
  \begin{center}
    \begin{tabular}{|c|c|c|} \hline
Process                                      & Rejection factor & Expected events  \\
\hline
e$^+$e$^-\to\omega\pi^0\to\pi^0\pi^0\gamma$  & 140 & 54$\pm$ 6 \\
$\phi\to\pi^0\pi^0\gamma$                    & 40 & 152$\pm$16 \\
$\phi\to\eta\gamma\to\gamma\gamma\gamma$     & 6$\times$10$^{4}$& 5$\pm$2 \\
$\phi\to\eta\gamma\to\pi^0\pi^0\pi^0\gamma$  & 2.5$\times$10$^{3}$& 98$\pm$10 \\
    \hline
    \end{tabular}
    \caption{Rejection factors for the background processes contributing to
      the five photon final state.
      The errors on the number of expected events include MC
      statistics and systematics from cross section uncertainties.}
    \label{effi}
  \end{center}
\end{table}

\begin{figure}
\begin{center}
\mbox{\epsfig{file=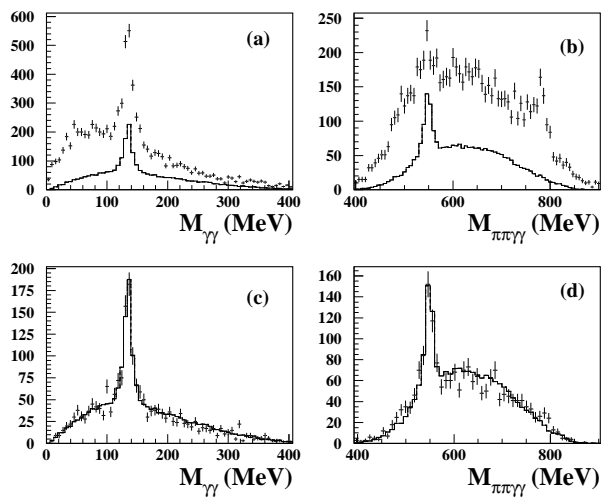,width=9cm}}
\end{center}
\caption{\small Distributions of the invariant mass of all
  $\gamma\gamma$
  and $\pi^+\pi^-\gamma\gamma$
  combinations for the events surviving fit1 ((a) and (b)) and fit2
  ((c) and (d)). The distributions are
  compared with the MC ones (histograms) for the signal only.}

\label{Figc1}
\end{figure}

\begin{figure}
\begin{center}
\mbox{\epsfig{file=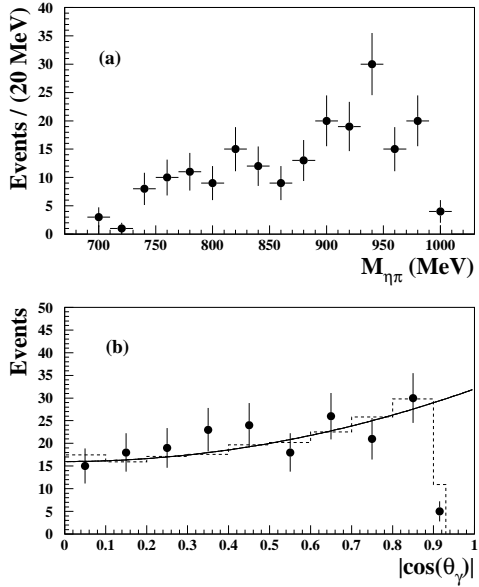,width=7cm}}
\end{center}
\caption{\small $\eta\to\pi^+\pi^-\pi^0$ sample. 
(a) Distribution of the $\eta\pi^0$ invariant mass
for the final sample of events; (b) distribution of
$|$cos$(\theta_{\gamma})|$ for the final sample of events. The dashed
histogram is the MC, and the curve
superimposed corresponds to a $(1+$ cos$^2(\theta_{\gamma}))$ dependence.}
\label{Figc2}
\end{figure}

We next consider the case of $\eta\pi^0\gamma$ in which  $\eta\rightarrow
\pi^+\pi^-\pi^0$.
This decay chain  gives a final state with 2 charged pions and 5 prompt
 photons.
This signature is unique among the
possible final states so that the
main background sources come from final states with similar topologies and
much larger branching ratios. The most significant backgrounds are:
\begin{itemize}
\item[-]
{ $\phi \rightarrow \eta\gamma$ with
$\eta\rightarrow \pi^+\pi^-\pi^0$ (2 tracks and 3 photons);}
\item[-]
{$e^+e^-\rightarrow \omega\pi^0$ with $\omega \rightarrow \pi^+\pi^-\pi^0$
(2 tracks and 4 photons); }
\item[-]
{ $\phi \rightarrow K_SK_L$ with
a prompt $K_L$
decay and a combination of $K_S\rightarrow \pi^+\pi^-$ and
$K_L\rightarrow \pi^0\pi^0\pi^0$ or $K_S \rightarrow \pi^0\pi^0$ and
$K_L\rightarrow \pi l \nu$ or $K_L\rightarrow \pi^+\pi^-\pi^0$ resulting in
2 tracks and 4 or 6 photons.}
\end{itemize}
The signal events are selected by requiring a vertex close to the interaction
region 
with two tracks of opposite charge, and five 
prompt photons with
 $E_{\gamma}>$10 MeV and $|$cos$\theta|<0.93$. The surviving events ($7.1\times
 10^3$ )
enter a two-step kinematic fit procedure:
a first fit (fit1) with energy/momentum
conservation at the vertex, and a second fit (fit2) with $\pi^0$ and $\eta$ 
mass constraints, where the combination resulting in the higher
$\chi^2$ probability is selected. The invariant mass distributions for
all $\gamma\gamma$ and $\pi^+\pi^-\gamma\gamma$ combinations are
shown in
Fig. \ref{Figc1}(a) and (b) for the events surviving fit1.
A clear $\eta$ signal already emerges
at this stage over a large combinatorial background due to
residual $\omega\pi^0$ and $K_SK_L$ events  (see the peak close to the
$\omega$ mass in Fig. \ref{Figc1}(b)).
After fit2 the residual background
is reduced to 4$\pm$4 events from $\omega\pi^0$ and less than 8 events 
from the other possible background modes. Fig. \ref{Figc1}(c) and (d) shows 
the $\gamma\gamma$ and $\pi^+\pi^-\gamma\gamma$ invariant mass 
distributions at this 
stage compared with the MC distributions for the signal normalized to 
the same number of events. The comparison shows good agreement. 
The distribution of $M_{\eta\pi}$ for the 197 events found 
is shown in Fig. \ref{Figc2}(a). The angular distribution of the radiated 
photon, Fig. \ref{Figc2}(b), agrees with (1+cos$^2\theta_{\gamma}$).

The efficiency (see Fig. \ref{effvsm}) as a function of $M_{\eta\pi}$ is 
found by using MC simulation. Both photon and 
track efficiencies are corrected using data control samples.
($K_S\rightarrow \pi^+\pi^-$ for tracking and $e^+e^-\gamma$ for $\gamma$'s). 
The overall efficiency, on average 19\%, is dominated 
by the probability for at least one of the seven particles to go out of 
acceptance.
The resolution in $M_{\eta\pi}$ is 4 MeV over the entire mass range. A non 
Gaussian tail of about 10\% is present, due to incorrect photon pairings.

The $\phi\rightarrow\eta\pi^0\gamma$ branching ratio is obtained independently 
for each sample, normalizing the number of events after background 
subtraction, $N-B$, to the $\phi$ cross section  $\sigma_{\phi}$, to the
selection efficiency $\epsilon$, and to the 
integrated luminosity $L$:
\begin{equation}
{\rm BR}(\phi\rightarrow\eta\pi^0\gamma)={N-B\over 
\epsilon\times {\rm BR}(\eta\to i)}\x 
{1\over\sigma_{\phi}\times L}.
\end{equation}
where $i$ = \gam\gam, \pic\po\ and BR(\et\to $i$)
is from ref. \cite{PDG}.
The luminosity is measured using large angle Bhabha scattering events.
The $\sigma_{\phi}$ measurement is obtained using the 
$\phi\to\eta\gamma\to\gamma\gamma\gamma$ decay \cite{KN_cs}.\\
\black

We obtain for the sample in which $\eta\to\gamma\gamma$ 
$$\eqalign{%
{\rm BR}(\f\to\et\po\gam)&=(8.51\pm0.51_{\rm stat}\pm0.57_{\rm
syst})\x10^{-5}\cr}$$
and for the sample in which $\eta\to\pi^+\pi^-\pi^0$
$$\eqalign{%
{\rm BR}(\f\to\et\po\gam)&=(7.96\pm0.60_{\rm stat}\pm0.40_{\rm
syst})\x10^{-5}\cr}$$
The two values are in agreement. Our results also agree with 
those from SND $(8.8\pm1.4\pm0.9)\times 10^{-5}$ \cite{snda0} and CMD-2 
$(9.0\pm2.4\pm1.0)\times 10^{-5}$\cite{cmd2a0}.
Contributions to the uncertainties are listed in Tab. \ref{Tabnc}.
\begin{table}
\begin{center}
\vskip 0.1 in
\begin{tabular}{|c|c|c|} \hline
  & Type 1 & Type 2 \\
\hline
 statistics  &  0.43  &  0.58  \\
 background subtraction & 0.28 & 0.15 \\
 efficiency & 0.51 & 0.30 \\
 $BR(\eta\to x)$ & 0.05 & 0.14 \\
 luminosity & 0.17 & 0.16 \\
 $\phi$ cross section & 0.17 & 0.16 \\
\hline
\end{tabular}
\caption{ Contributions to the uncertainties in
  $BR(\phi\rightarrow\eta\pi^0\gamma)$ measurement ($10^{-5}$ units).}
\label{Tabnc}
\end{center}
\end{table}

We estimate the contribution of \f\to\ao\gam\ from a simultaneous fit of the 
two $M_{\eta\pi}$ spectra. Two contributions are considered in the fit:
 (a) \f\to$\rho^0$\po, $\rho^0$\to\et\gam\ and (b) \f\to\ao\gam, 
\ao\to\et\po. The contribution from the continuum process 
 $e^+e^-\rightarrow\omega\pi^0$, $\omega\to\eta\gamma$ 
is negligible due to the low cross section. The $M_{\eta\pi}$ spectrum 
for (a) is taken from VMD 
calculations \cite{Achasov,Bramon}. For (b) we use the formulation of 
ref.\cite{Achasov89} based on $\phi$ coupling to a charged kaon loop:
\begin{equation}
{\dif\Gamma(\f\to a_0\gam\to\et\po\gam)\over\dif M_{\eta\pi}}=
{2M^2_{\eta\pi}\over\pi}{\Gamma_{\f a_0\gam}\Gamma_{\ao\et\po} 
\over|D_{a_0}|^2}
\label{scalare}
\end{equation}
where $\Gamma_{\phi{a_0}\gamma}$ is related to the 
coupling $g^2_{a_0K^+K^-}/4\pi$, $\Gamma_{a_0\eta\pi^0}$ to the 
coupling $g^2_{a_0\eta\pi^0}/4\pi$, 
and $D_{a_0}$ is the inverse $a_0$ propagator including finite width 
corrections.
The model assumes point-like couplings.

\begin{figure}
\begin{center}
\mbox{\epsfig{file=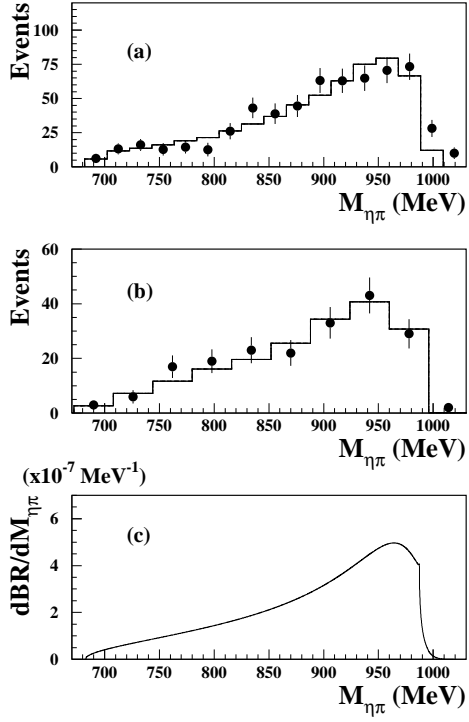,width=7cm}}
\end{center}
\caption{\small Result of the combined fit: (a) comparison of data
(exp. points) vs. fit (histogram) for
  (a) type 1 (b) type 2 samples and (c) plot of the theoretical curve for the
  $a_0$ contribution with the parameters extracted from the fit. }
\label{fit}
\end{figure}

The efficiency and resolution functions, including distortions from incorrect
photon pairings are folded into the theoretical distribution. 
The experimental spectra after background 
subtraction (Fig.\ref{fit}) 
are simultaneously fitted, setting $M_{\ao}$=984.8 MeV, from 
ref \cite{PDG}. The free parameters of the fit are
the branching ratio
for contribution (a) and the two coupling constants above.
We find BR(\f\to$\rho^0$\po, $\rho^0$\to\et\gam)=$(0.5\pm0.5)\times 10^{-5}$, 
$g^2_{\ao K^+K^-}/4\pi=(0.40\pm0.04)$ GeV$^2$, and
$g_{\ao\eta\pi}/g_{\ao K^+K^-}=1.35\pm0.09$. The contribution from 
\f\to\ao\gam\ is dominant and that from $\rho^0$\po\ is consistent with zero, 
in agreement with VMD
calculations \cite{Bramon}.
By integration we find:
\begin{equation}
 BR(\phi\rightarrow a_0\gamma,a_0\rightarrow\eta\pi^0)=(7.4\pm0.7)\times
 10^{-5}
\end{equation}
The fit is good, with $\chi^2$/ndf=27.2/25. 
Fig. \ref{fit} shows the fit and the $a_0$ contribution.
Interference between (a) and (b) has been considered \cite{Achasov} giving 
no significant effect in the fit results.
 
The result for $g_{\ao\eta\pi}/g_{\ao K^+K^-}$ can be compared
with the value $1.05\pm0.06$ given by experiment BNL852 \cite{BNL852} and 
with the Crystal Barrel results (1.07
or 0.93 depending on the analysis \cite{Crystal}).
It can also be compared with
expectations based on the standard $q\bar q$ model (1.51 according to
\cite{Achasov89}) and on the $q\bar qq\bar q$ model (0.93 according to 
\cite{4quarks}).

Finally,
the results are combined with those obtained
in the analysis of $\phi\rightarrow \pi^0\pi^0\gamma$
\cite{kloef0} done on the same data sample:
\begin{itemize}
\item[-]{
$BR(\phi\rightarrow f_0\gamma\rightarrow\pi^0\pi^0\gamma)=(1.49\pm
0.07)\times 10^{-4}$;}
\item[-]{$g^2_{f_0K^+K^-}/4\pi=2.79\pm0.12$ GeV$^2$. }
\end{itemize}
Multiplying the branching ratio above by a factor of 3 to 
account for $f_0\rightarrow \pi^+\pi^-$ decays BR(\f\to$f_0$\gam) is then 
$(4.47\pm 0.21)\times 10^{-4}$, and the ratio of the two branching ratios is\\
\parbox{\textwidth}{\begin{equation}
R_{BR}={{\rm BR}(\phi\rightarrow f_0\gamma)\over
 {\rm BR}(\phi\rightarrow a_0\gamma)} = 6.1\pm0.6
\end{equation}}\\
The ratio of the two couplings to the $KK$ system is
\begin{equation}
R_{g^2}={{g^2_{f_0KK}}\over {g^2_{a_0KK}} } = 7.0\pm0.7
\end{equation}
These ratios are useful for shedding light on the possible sizes and mixing or
other structure questions regarding these mesons \cite{close,delusion}.

\ack
We thank the DA$\Phi$NE team for their efforts in maintaining low
background running 
conditions and their collaboration during all
data-taking. We also thank F. Fortugno for 
his efforts in ensuring
good operations of the KLOE computing facilities. This work was
supported in part by DOE grant DE-FG-02-97ER41027; by 
EURODAPHNE, contract FMRX-CT98-0169; 
by the German Federal Ministry of Education and Research (BMBF) contract 
06-KA-957; 
by Graduiertenkolleg 'H.E. Phys.and Part. Astrophys.' of 
Deutsche Forschungsgemeinschaft, 
Contract No. GK 742;
by INTAS, contracts 96-624, 99-37; and by TARI, contract 
HPRI-CT-1999-00088.




\end{document}